# A Dynamic Dark Information Energy Consistent with *Planck* Data

Michael Paul Gough

Department of Engineering and Design, University of Sussex, Brighton, BN1 9QT, UK

E-Mail: m.p.gough@sussex.ac.uk

**Abstract:** The 2013 cosmology results from the European Space Agency *Planck* spacecraft provide new limits to the dark energy equation of state parameter. Here we show that Holographic Dark Information Energy (HDIE), a dynamic dark energy model, achieves an optimal fit to the published datasets where *Planck* data is combined with other astrophysical measurements. HDIE uses Landauer's principle to account for dark energy by the energy equivalent of information, or entropy, of stellar heated gas and dust. Combining Landauer's principle with the Holographic principle yields an equation of state parameter determined solely by star formation history, effectively solving the 'cosmic coincidence problem'. While HDIE mimics a cosmological constant at low red-shifts, $z<1$, the small difference from a cosmological constant expected at higher red-shifts will only be resolved by the next generation of dark energy instrumentation. The HDIE model is shown to provide a viable alternative to the main cosmological constant/vacuum energy and scalar field/quintessence explanations.

## 1. Introduction

Measurements of type 1a supernova [1,2] show that a dark energy of unknown origin has caused an acceleration of the universe expansion in recent times. This result has since been confirmed by several independent measurements [3-7]. Favoured explanations for this dark energy include a cosmological constant, or vacuum energy, and some form of scalar field, or quintessence [8,9]. Vacuum energy, the simplest explanation and mathematically equivalent to a cosmological constant, is predicted by quantum field theories to have 120 orders of magnitude energy density, or 30 orders of mass scale, greater than that observed, and incompatible with our existence [8,10,11]. Scalar fields require fine tuning, both to account for the observed value, and to provide a constant dark energy density [8,9]. Neither the cosmological constant nor scalar fields seem able to solve the cosmic coincidence problem.

Here we show the Holographic Dark Information Energy (HDIE) model [12,13] provides a reasonable account of dark energy, and can solve the cosmic coincidence problem, just by taking a simple phenomenological approach. HDIE proposes that dark energy is the energy equivalent of information, or entropy, associated with stellar heated gas and dust. This present work shows that the HDIE model provides a good fit to the dark energy values derived from the recently released results of the *Planck* mission [14] and compares favourably with the two main dark energy theories.

## 2. Review of the HDIE model

*2.1 Two Principles*

The HDIE model applies two foundational principles from information theory:

<u>Landauer's Principle: Information-energy equivalence</u>

Landauer's principle [15-18] provides a minimum information-energy equivalence of $k_BT\ln2$ per bit, where $k$ is the Boltzmann constant and $T$ is temperature. Landauer's principle effectively resolves the paradox of Maxwell's demon [19] and has now been verified by experiments [20,21]. A single colloidal particle was trapped in a modulated double-well potential [21] to form a one bit memory, replicating the situation previously considered theoretically by Landauer [16]. In the limit, the mean dissipated heat from information erasure was found to saturate at $k_BT\ln2$ per bit, demonstrating that information is indeed physical [15], and closely linked to thermodynamics. Clearly, when the same degrees of freedom are being considered, information and entropy are identical with 1bit= *ln2* nats.

It is worth noting that information erasure regularly occurs in normal computer operation, every time that a memory location is overwritten with a new value. However, the Landauer heat, $k_BT\ln2$, generated by erasing each bit is miniscule, $\leq 10^{-10}$ of the normal electronic energy dissipation, $CV^2/2$, produced when erasing a bit by discharging the charge on the capacitance, $C$, of the gate of a CMOS memory cell operating at a supply voltage, $V$ [13, 22].

While too small to affect our electronics for some years to come, the Landauer information energy is shown below to be making a significant contribution to the universe energy balance. Here we are concerned with the energy equivalence of information in the universe. From a cosmology point of view it is more important to assess the energy represented by that information, rather than to identify information erasure processes generating heat.

<u>Holographic Principle: Information scales with bounding area</u>

The Holographic principle [23-25] asserts that the number of degrees of freedom in any region of space is proportional to the area of its boundary, rather than to its volume. The Holographic principle has been proposed [25] as a general principle, not just limited to black holes at the maximum entropy holographic bound. Black holes can be considered as having their information packed at the holographic bound with each bit taking up one unit area of (Planck length)$^2$ on the black hole event horizon. However, Table 1 clearly shows that the universe information total is many orders of magnitude below the universe's holographic bound, $\sim 10^{124}$. In contrast to a black hole, the universe baryon bit totals $\sim 10^{86}$ correspond to areas per bit on the universe event horizon of the order of units of (Fermi length)$^2$. While the components of black holes are ultimately compacted to Planck lengths, $1.6 \times 10^{-35}$m, representing the smallest physically significant distance, baryons in the universe move more freely and are better described at nuclear scale distances, characterised by the Fermi length, $10^{-15}$m.

The Holographic Principle is supported by string theory with a well-known quantum theory example. The 'Maldacena duality', or 'anti de-Sitter / conformal field theory' (AdS/CFT) [26], permits one particular multi-dimensional space with gravity to be translated into another with one less dimension without gravity, equivalent to a holographic translation. This result is now strongly supported by recent theoretical work on another example [27,28], but remains only proven for some cases of multi-dimensional space. Such a holographic translation for the specific case of the universe that we live in remains to be proved.

While Landauer's Principle has been experimentally proven [21], the Holographic Principle is an attractive conjecture that, by its very nature, has turned out to be difficult to verify [29]. The Holographic Principle is required to explain the time history of HDIE and thus represents the main weakness in the HDIE model. Fortunately, we show below that HDIE can account quantitatively for the present dark energy density value just from Landauer's Principle, without calling on the Holographic Principle. This encourages us to proceed with HDIE, since the difficulty other dark energy theories have in accounting for today's value leads to the 'cosmic coincidence problem' [8-11].

*2.2 Universe information energy contributions*

Table 1. lists the relevant components of the universe, together with estimates of the quantity of information, $N$, associated with them [30, 31], representative temperatures, $T$, and the resulting information energy, $N\,k_B\,T\,ln2$, for each component.

| | | **Information, N bits** | **Temperature, T °K** | **Information Energy $N\,k_B\,T\,ln2$, Joules** |
|---|---|---|---|---|
| **Relics of Big Bang** | CMB photons | $10^{88} - 2 \times 10^{89}$ | 2.7 | $3 \times 10^{65} - 6 \times 10^{66}$ |
| | Relic neutrinos | $10^{88} - 5 \times 10^{89}$ | 2 | $2 \times 10^{65} - 10^{67}$ |
| | Relic gravitons | $10^{86} - 6 \times 10^{87}$ | ~1? | $10^{63} - 6 \times 10^{64}$ |
| **Dark matter** | Cold dark matter | $\sim 2 \times 10^{88}$ | $<10^2$ ? | $< 10^{67}$ |
| **Star formation** | $10^{22}$ stars | $10^{79} - 10^{81}$ | $\sim 10^7$ | $10^{63} - 10^{65}$ |
| | Stellar heated gas and dust | $\sim 10^{86}$ | $\sim 10^6 - 10^7$ | $\sim 10^{69} - 10^{70}$ |
| **Black Holes** | Stellar sized BH | $10^{97} - 6 \times 10^{97}$ | $\sim 10^{-7}$ | $10^{67} - 6 \times 10^{67}$ |
| | Super massive BH | $10^{102} - 3 \times 10^{104}$ | $\sim 10^{-14}$ | $10^{65} - 3 \times 10^{67}$ |
| **Universe** | Holographic bound | $\sim 10^{124}$ | - | - |

**Table 1.** Universe information content, temperature, and information energy contributions.

Although there is considerable uncertainty in the above information bit numbers, stellar heated gas and dust seem to provide the majority of the information energy of the universe, followed by black hole information energy, at $<10^{-2}$ of the stellar heated gas and dust value. Note that, from the universe's point of view, the black hole "no hair theorem" [32] implies that the information represented by each black hole is similar to that of just one single fundamental particle with only three relevant parameters: mass, charge, and spin. Furthermore, although information is thought to return to the universe through evaporation from black holes [33], such evaporation will occur over such long timescales, $\sim 10^{67}$ years, as to have negligible effect on the present. For these reasons, we only consider here the information energy of stellar heated gas and dust.

*2.3 Stellar heated gas and dust*

HDIE can easily provide [12,13] today $N\, k_B\, T\, ln2$ ~$10^{70}$J of energy, equivalent to the energy of the mass of the observable universe (~$10^{53}$kg), and thus of a similar order to the present dark energy value. This value derives simply from order of magnitude estimates of the entropy of stellar heated gas and dust, $N$~$10^{86}$ bits [30,31] with typical stellar baryon temperatures, $T$~$10^{7}$K.

HDIE total energy varies over time in proportion to the product of average baryon temperature, $T$, and total information content, $N$. A survey of nine datasets of integrated stellar density measurements [34-42] are plotted in Fig. 1 (adapted from Fig. 1a of reference [13] ) against the cosmological scale factor, $a$, defined in terms of redshift, $z$, by $a = 1/(1+z)$.

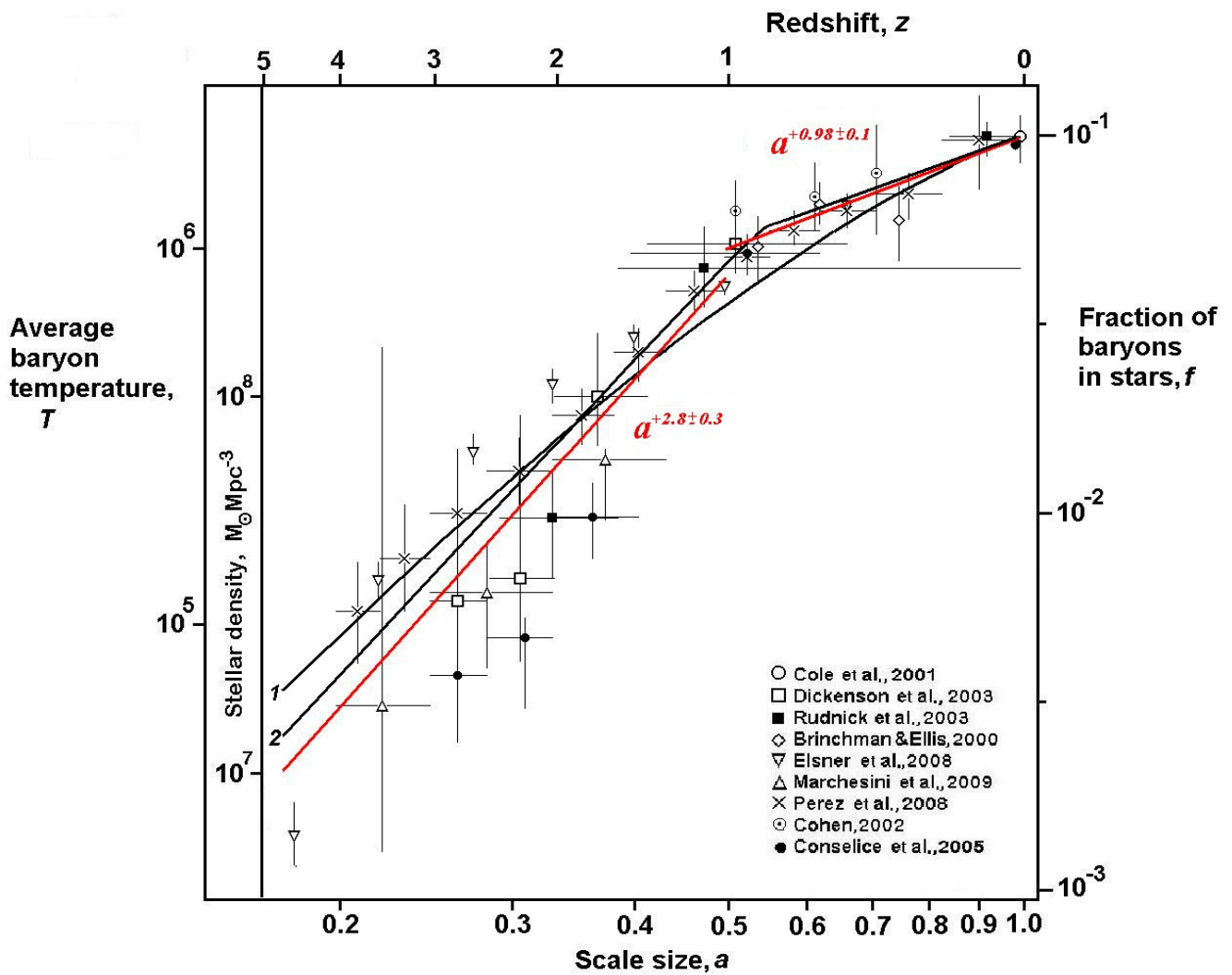

**Figure 1.** A survey of integrated stellar density measurements with corresponding fraction of baryons in stars and average baryon temperature. Nine datasets are plotted as solar masses per cubic mega parsec, with the red lines illustrating the power law fits to data either side of $z$=1. Continuous black line *1*: for an HDIE equation of state parameter of the form $w(a)= w_o + (1 - a)w_a$, with $w_o= -1$ and $w_a= -0.6$. Continuous black line *2*: for an HDIE equation of state parameter of the form $w(a)= w_o + (w_a / ( 1+ \exp((a-a_t)/a_w)))$, with $w_o= -1$, $w_a= -0.6$, width of transition, $a_w= 0.05$, and transition, $a_t$, at $z= 0.9$.

These datasets show a distinct change in power law near $z\sim1$. When least squares curve fitting is applied to the logarithmic values, the fraction of baryons in stars, and therefore $T$, is found to vary as $a^{+2.8\pm0.3}$ at early times, $z>1$, changing to a lower rate of $a^{+0.98\pm0.10}$ in recent times, $z<1$, illustrated by the red lines in Fig 1. Assuming the information content, $N$, varies according to the holographic principle, proportional to the bounding area as $N \propto a^2$, the combined $NT$ variation leads to a total HDIE dark energy varying as $a^{+4.8\pm0.3}$ for $z>1$, and $a^{+2.98\pm0.10}$ for $z<1$. This corresponds to energy densities that vary as $a^{+1.8\pm0.3}$ for $z>1$, but effectively constant for $z<1$ varying as $a^{-0.02\pm0.10}$, causing HDIE to mimic a cosmological constant in recent times.

The HDIE equation of state parameter, $w$, the ratio of pressure to energy per unit volume and defined by energy densities varying as $a^{-3(1+w)}$, is therefore determined solely by star formation history. HDIE energy densities of $a^{+1.8\pm0.3}$ for $z>1$, and $a^{-0.02\pm0.10}$ for $z<1$, then correspond to HDIE equation of state parameter ranges $-1.5>w>-1.7$ for $z>1$, and, in recent times, $z<1$, to $-0.96>w>-1.03$.

Thus HDIE can account quantitatively for the two most important properties of dark energy: the present energy value ($\sim 10^{70}$ J) and the recent period with an equation of state, $w \sim -1$.

### 3. *Planck* dark energy measurements.

*3.1 Parameterisation of w(a)*

A key test for dark energy theories is to compare observed and predicted variations of the dark energy equation of state parameter, $w(a)$, over time. It is conventional to use a parameterisation, where the present value is denoted by $w_o$, and the early value ($z\gg1$) denoted by $w_o + w_a$. This allows for the possibility of either a cosmological constant ($w_o = -1$ and $w_a = 0$) or some form of dynamic dark energy ($w_a \neq 0$). The commonest form is given by $w(a) = w_o + (1 - a) w_a$ [43]. Line *1* in Fig. 1. illustrates this form of parameterisation fitting the data with the specific values $w_o = -1$ and $w_a = -0.6$, for the expected HDIE equation of state parameter values: $w = -1.6$ for $z\gg1$; and $w = -1.0$ for $z<1$. However, we can see from Fig. 1. that this form of parameterisation provides a much slower transition than we expect for HDIE from the star formation data. Those measurements lead us to expect a more abrupt transition. A preferred parameterisation is given by $w(a) = w_o + (w_a /( 1 + \exp((a-a_t)/a_w)))$. This four parameter description has been used previously [44] and is illustrated by line *2* in Fig. 1, where again $w_o = -1$, $w_a = -0.6$, but now with a narrower width of transition, $a_w = 0.05$, and with a location of transition, $a_t$, corresponding to $z=0.9$. But, since the introduction of two extra variables further complicates data fitting, and the simpler two parameter description has been already applied in the published *Planck* data analysis [14], we continue here using that more usual form.

*3.2 Planck cosmological parameter data*

The latest 2013 cosmological parameter results [14] from the European Space Agency Planck spacecraft uses *Planck* Cosmic Microwave Background (CMB) data combined with other astrophysical datasets to provide information on the dark energy equation of state parameter.

In order to place limits on the dark energy equation of state parameter, *Planck* CMB data, combined with WMAP CMB polarization data [45], has to be further combined with at least one other, non

CMB, type of astrophysical measurement. Several combinations were considered: *Planck*+WMAP+BAO, combining with baryon acoustic oscillation data; *Planck*+WMAP+Union2.1, combining with a group of 580 type 1a supernovae [46]; and *Planck*+WMAP+SNLS combining with a group of 473 type 1a supernovae [47]. The dark energy equation of state parameter was assumed to take the usual two parameter form $w(a) = w_o + (1 - a)w_a$. The 2D marginalised posterior distributions for $w_o$ and $w_a$ are plotted for three *Planck* data combinations in Fig. 2, adapted from Fig. 36 of reference [14].

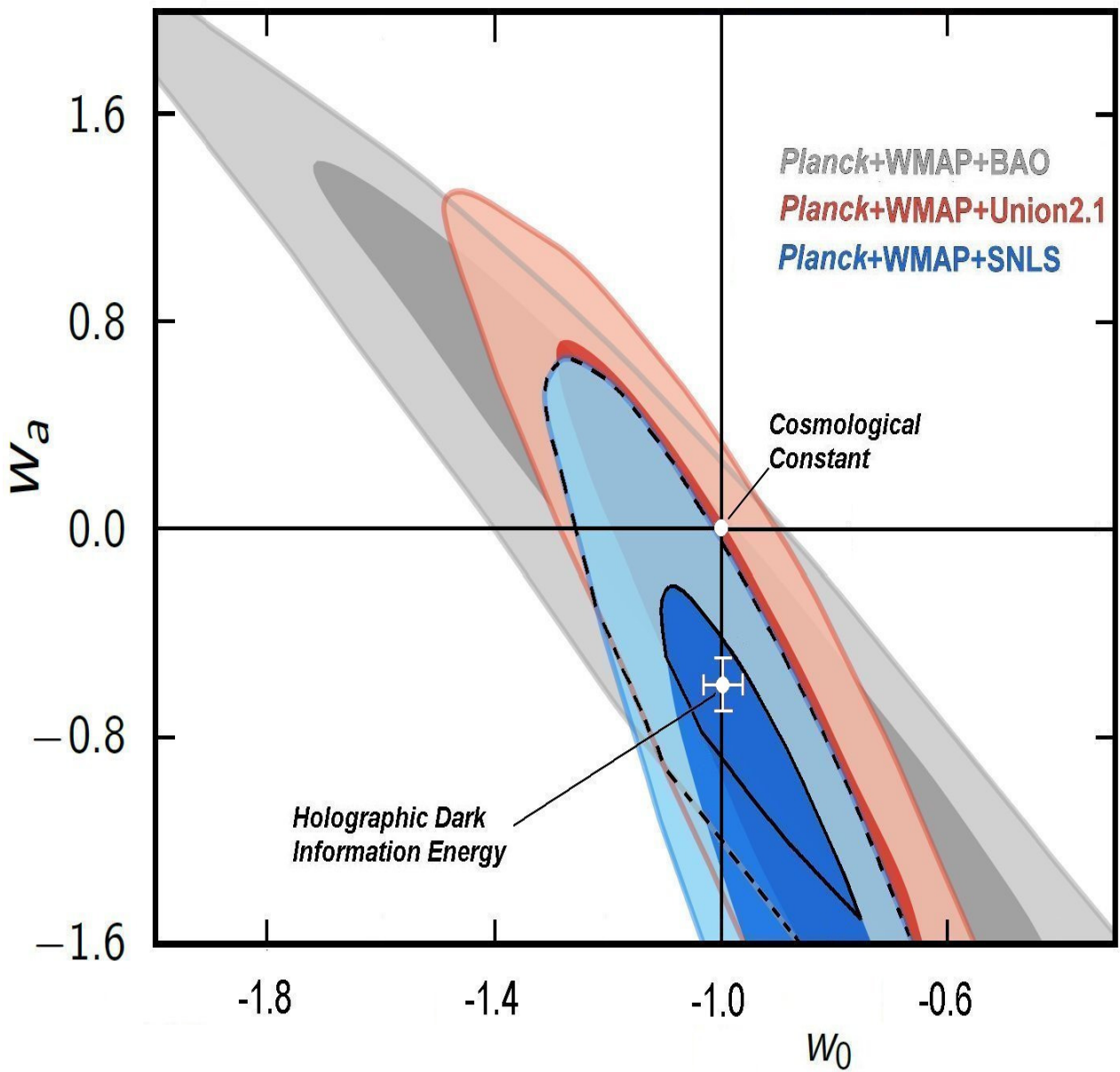

**Figure 2.** Combined *Planck* datasets for the dark energy equation of state parameter of the form: $w_o + (1 - a) w_a$, where $a$ is the cosmological scale factor (after fig. 36 of reference [14] ). 2D marginalised posterior distributions are shown by the 68% and 95% likelihood contours for the three *Planck* data combinations discussed in the text. The areas bounded by the black dashed line and the black continuous line correspond to the 95% and 68% likelihoods, respectively, that are common to all three dataset combinations. HDIE error bars are set by the 1σ errors in the stellar density measurement fits to power laws shown in Fig 1.

*3.3 Planck data comparison with cosmological constant and HDIE models*

The cosmological constant can be seen to lie just inside the 68% likelihood contours of the *Planck*+WMAP+BAO and *Planck*+WMAP+Union2.1 data combinations, but just outside of even the 95% likelihood contour of the *Planck*+WMAP+SNLS data combination. Both the *Planck*+WMAP+SNLS and another, fourth, combination of *Planck* data with recent measurements [48] of the Hubble constant, show dark energy to be dynamic at the 2σ level, while the *Planck*+WMAP+BAO and *Planck*+WMAP+Union2.1 combinations are compatible with a cosmological constant. Since 2σ is not a difference of very high significance, the authors of reference [14] were still inclined to favour a cosmological constant. However, they noted that all four *Planck* combinations would be reconciled by a dynamic dark energy exhibiting $w<-1$ at earlier times, corresponding to $w_a<0$.

The HDIE equation of state parameter ranges, $-1.5>w>-1.7$ for $z>1$, and $-0.96>w>-1.03$ for $z<1$, correspond to $w_o$, $w_a$ values: $-0.96>w_o>-1.03$ and $-0.5>w_a>-0.7$, shown in Fig.2 alongside the *Planck* data. For comparison with stellar density survey data and power law fits to that data, the continuous line *1* in Fig 1. is the temperature profile required to produce an HDIE equation of state parameter of the same form as used in the *Planck* analysis, $w(a)= w_o + (1 - a) w_a$, with the specific values $w_0= -1$ and $w_a= -0.6$.

Then, in contrast to the location of the cosmological constant, HDIE is found to lie centrally within the continuous black line enclosed region of Fig.2, inside the 68% likelihood contour of all three data combinations. The centre of the common dataset 68% region of Fig. 2 at the value $w_o = -1$ is found to be located close to $w_a= -0.6$, corresponding to a dark energy density that increases as $a^{+1.8}$ for $z>>1$, as expected for HDIE from the stellar formation data of Fig.1.

We saw in Fig.1 that the simplest two variable parameterisation, $w(a)=w_o+(1-a)w_a$, does not sufficiently represent HDIE. Recent Baryon Oscillation Spectroscopic Survey (BOSS) measurements with an accuracy of 1% are found to be fully consistent with a cosmological constant type of behaviour in the redshift range $0.2<z<0.7$ [49]. However, the Hubble parameter of a dynamic dark energy described by the simple two variable parameterisation with $w_o=-1$, and $w_a =-0.6$ differs from a cosmological constant by 11% at $z=0.7$. In order to reflect the fairly abrupt transition expected near $z\sim 1$, and maintain the identical nature of HDIE to a cosmological constant for $0<z<1$, it is clearly better to use the four parameter description, $w(a)=w_o+(w_a/( 1 + \exp((a-a_t)/a_w)))$ in future data analysis.

*3.4 Hubble parameter measurements.*

The mass density, falling steeply as $a^{-3}$, dominated the energy contributions at earlier times. This makes it very difficult to distinguish between HDIE and a cosmological constant at $z>1$ where the only measureable difference is expected to be found. Ideally, instead of integrating data by applying a parameterisation of the equation of state parameter, we should have the ability to measure the Hubble parameter, $H(a)$, at very high resolution over a range of redshifts, $z>1$.

A survey of recent Hubble parameter, $H(a)$, and Hubble constant, $H_0$, measurements [48-56] is shown in Fig.3. These measurements are plotted as $H(a)/(1+z)$, illustrating the change from deceleration to acceleration. For comparison, we also plot as red continuous lines the variations expected for HDIE, and for a cosmological constant for the specific case of the *Planck* consortium derived cosmological parameters[14]: 68.5±1.7% dark energy and Hubble constant $H_0$=67.3±1.2 km/sec/Mpc. As one of the *Planck* detector bands at 217GHz is thought to have introduced tension between the *Planck* results and previous astronomical measurements [14, 50], *Planck* data has been re-analysed with less emphasis on this band. The blue lines correspond to the values 69.8±1.5% dark energy and $H_0$=68.0±1.0 km/sec/Mpc derived from the re-analysed data [50].

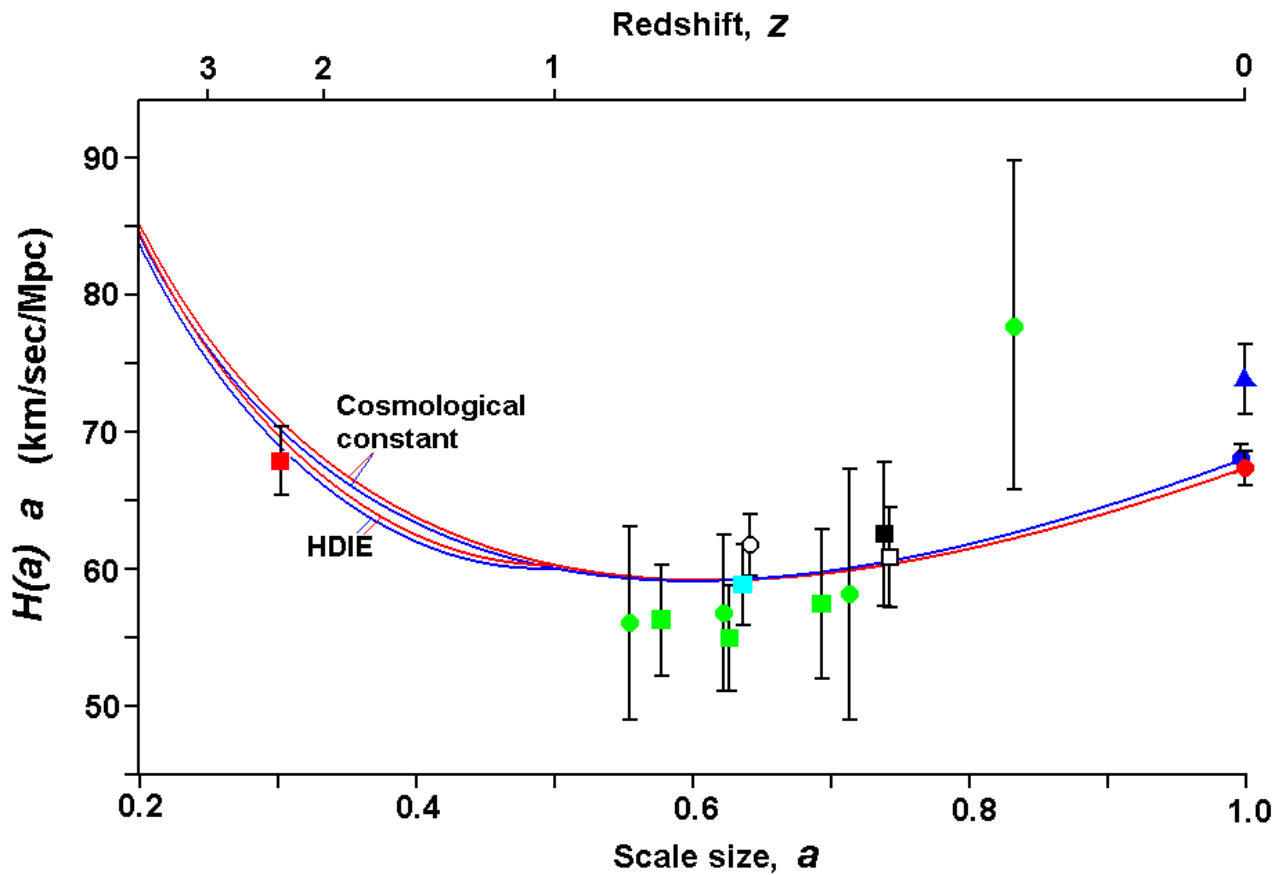

**Figure 3.** Measurements of the Hubble Parameter $H(a)$ plotted as $H(a)/(1+z)$, or $H(a)$ $a$, against universe scale size, $a$, illustrating the change from deceleration to acceleration. Hubble constant, $H_0$, measurements: blue triangle, Reiss [48], red circle *Planck* consortium [14], and blue circle, Spergel re-analysed *Planck* data[50]. Hubble parameter, $H(a)$ measurements: open circle, Anderson [49]; red square, Busca [51]; green circles, Blake [52]; green squares, Blake [53]; cyan square, Reid [54]; filled black square, Xu [55] and open square, Chuang [56]. Variation expected for the cosmological constant and for the HDIE dark energy model shown as continuous lines: red lines assuming *Planck* consortium values for $H_0$ (67.3±1.2 km/sec/Mpc) and dark energy (68.5±1.7%) [14]; and blue lines assuming Spergel [50] re-analysed *Planck* data values, $H_0$ (68.0±1.0 km/sec/Mpc) and dark energy (69.8±1.5%).

It is clear from Fig.3. that at present we are unable to resolve the relatively subtle difference between a cosmological constant and the HDIE model, because of the current paucity of measurements z>1, and because of the limited resolutions of existing $H(a)$ instruments. However, this small difference should be resolved by future instruments that achieve resolutions of $\Delta H/H \sim 1\%$ in the range 1<z<2. Such measurements should be made by around the year 2020, when the next generation of space and ground-based instruments are operating: Euclid [57]; WFIRST [58]; BigBOSS [59]; LSST [60]; and Dark Energy Survey [61].

## 4. Cosmic Coincidence Problem(s)

*4.1 The 'Cosmic Coincidence Problem'*

The well known 'cosmic coincidence problem' asks why the dark energy density has a similar value to the present mass energy density, so that the accelerated expansion only started in the recent past? If the dark energy value had been stronger, dominating the universe at an earlier time, the faster expansion would have stopped stars being formed and we would not be here to observe. Neither vacuum energy nor quintessence seem able to explain the present dark energy density. However, the HDIE explanation readily accounts for the present energy density to order of magnitude accuracy with an explanation directly dependant on the integrated star formation history. Significant star formation was required both for HDIE generated dark energy and for our own existence. Should HDIE be proven to be the source of dark energy, this would effectively remove the 'cosmic coincidence problem'.

Our information based approach then identifies three more coincidences below that can also be resolved by, and thus provide support for, the HDIE explanation for dark energy.

*4.2 Recent integrated star formation rate ~ $a^{+1}$*

Why is the observed $a^{+0.98\pm0.10}$ integrated rate of star formation since $z\sim1$ just the right value to yield an HDIE equation of state parameter, $-0.96>w>-1.03$, closely centred around $w=-1$ for a near constant dark energy density? The acceleration caused by dark energy limits matter-density perturbations which reduces the growth of structure [9]. The distinct reduction in star formation rate after $z\sim1.0$ has been previously attributed to the onset of acceleration, with the subsequent faster expansion acting to reduce the star formation rate [62]. It has been suggested [12,13] that the transition in star formation rate, to one centred around the specific $a^{+1}$ gradient providing constant HDIE energy density, results from feedback. The earlier, steeper $a^{+2.8\pm0.3}$ rate of increase in star formation, providing the $a^{+1.8}$ energy density increase, could not continue after $z\sim0.9$. This would have lead to even greater HDIE energy density with higher acceleration which in turn would have drastically reduced the rate of star formation to limit HDIE. Once HDIE initiated acceleration, feedback between acceleration and star formation resulted in a balance with a natural preference for a constant HDIE dark energy density, with $T \alpha\ a^{+1}$, at a density value comparable to the matter energy density at the time acceleration started.

*4.3 Universe's algorithmic entropy*

A simple Gedanken experiment [13] to estimate the algorithmic information/entropy of the universe provides a further information related coincidence. We expect that the algorithmic information of the universe, or information required to simulate the universe on a hypothetical super computer, should never decrease as that would imply some form of decrease in conventional thermodynamic entropy. However, estimates of the information needed as input to that hypothetical simulation showed that the recent acceleration of universe expansion was needed to ensure that the universe's algorithmic information did not decrease due to increasing star formation. The decrease in algorithmic information due to the reduction in dimensions needed to describe the 10% of baryons forming stars is found to be exactly countered by the increasing algorithmic information of the remaining 90% of baryons described by the faster increasing dimensions of the universe undergoing an accelerating expansion. This result further supports an information based dark energy explanation, and, moreover, is also consistent with a feedback between star formation and the accelerating expansion.

*4.4 Characteristic energy*

For our final coincidence we note that the energy equivalent of each bit of information associated with the 90% of universe baryons not involved in star formation is defined identically to, and has the same value as, the characteristic energy of the cosmological constant. Today, that bit energy value is $k_B T ln2$ ~$3 \times 10^{-3}$eV [13,63,64], corresponding to $T$~30K, typical of background dust temperatures[65]. While previously this low value for the characteristic energy was considered too small to relate to any interesting particle physics [66], the information based approach can explain this value as the energy equivalence of a bit of information. Note, however, that HDIE depends on the other 10% of baryons that are involved in star formation with higher temperatures ~$10^7$K, corresponding to equivalent bit energies, ~$10^2$eV.

**5. Comparison of HDIE with the main dark energy theories.**

The two main dark energy theories, a cosmological constant or vacuum energy, and scalar fields or quintessence, have been reviewed previously [8-11]. Here we compare HDIE with these two classes of theories to ascertain how well they manage to account for the observed features of dark energy.

Ideally, theories that aim to explain dark energy should satisfy the following requirements:

a) <u>Account for the observed constant dark energy density, $w$=-1, in the recent period, $z$<1.</u> Scalar field theories typically generate equation of state parameters in the relatively wide range -1<$w$<+1 [8,9,10], requiring much fine tuning to achieve the specific value $w$=-1. By definition, a cosmological constant, or vacuum energy, provides a constant dark energy density, $w$=-1. At low redshifts, $z$<1, HDIE also directly provides a near constant dark energy density, with $w$ restricted to the very narrow range: -0.96>$w$>-1.03. This requirement is clearly satisfied by both the cosmological constant and HDIE, while scalar fields experience considerable difficulty in meeting the requirement.

b) <u>Account for today's dark energy density value.</u> There is no underlying physics for a cosmological constant as such, but the quantum field theories for vacuum energy predict a value 30 orders of mass scale greater than the observed dark energy density [8-11]. Moreover, a zero valued vacuum energy density would be easier to explain by theory than the observed low dark energy density value [11]. Scalar field theories again require much fine tuning to explain the observed dark energy density value [8,9]. In comparison to the two main theories, HDIE directly provides a value of similar order of magnitude to that observed, simply from estimates of $N$ and $T$, using experimentally proven Landauer's principle [21]. Note again that HDIE achieves this requirement without recourse to the unproven Holographic Principle.

<u>c) Agree with *Planck* dataset combinations.</u> Two *Planck* datasets agree with a cosmological constant or vacuum energy ($w_o$= -1 and $w_a$= 0) but the other two *Planck* datasets support a dynamic dark energy ($w_a$<0) at the 2σ level [14]. Fig.2. shows that, while the cosmological constant lies just inside the 68% likelihood contours of two data combinations, it is located just outside of even the 95% likelihood contour of the third data combination. However, the 2σ level difference is clearly too small to exclude a cosmological constant explanation. Scalar field models are dynamic by their nature, but predict a wide range of values for $w_o$ and $w_a$, with little particular preference for the narrow range deduced from the *Planck* observations. Many scalar field models predict an energy density falling towards a constant vacuum energy density, $w_a$>0 [8,9,11], whereas Fig.2. shows that *Planck* data favours increasing dark energy densities at earlier times, $w_a$<0. We note that HDIE provides a $w_o$,$w_a$ data point ideally located in the centre of the 68% likelihood area common to all *Planck* datasets (see section 3.3 above).

<u>d) Resolve the 'Cosmic Coincidence' problem.</u> The present dark energy density is a similar order of magnitude to the present energy density of matter. The 'cosmic coincidence problem' exists because neither the cosmological constant / vacuum energy, nor scalar field / quintessence explanations are able to provide a strong argument in support of the dark energy density value observed today. In contrast, HDIE solves the 'cosmic coincidence problem' by successfully accounting for the present dark energy density value with a dark energy directly driven by star formation (see sections 4.1 & 4.2 above).

The ability of HDIE and the two main dark energy theories to satisfy the above requirements can then be summarized in Table 1.

| Dark energy theory requirement | Cosmological Const. / vacuum energy | Scalar Fields / quintessence | HDIE |
|---|---|---|---|
| a) Provide a constant energy density in recent times, $w$=-1 at $z$<1 | By definition constant, $w$=-1 | Wide range, -1<$w$<+1 | Near constant, -0.96>$w$>-1.03 |
| b) Account quantitatively for today's dark energy density value | Many orders of magnitude different. | Only by much fine tuning | Yes, directly ~$10^{70}$ J |
| c) Consistent with *Planck* $w_o$,$w_a$ data | Reasonable agreement | Not specific | Full agreement |
| d) Solve 'Cosmic Coincidence problem' | No | No | Yes |

**Table 1.** HDIE model compared with the two main dark energy theories.

As HDIE is driven by star forming regions, we can expect that the HDIE contribution to dark energy will differ between matter dense star forming regions and the very low densities of cosmic voids. Recent advances in gravitational lensing techniques have enabled the identification of a diminutive lensing signal, or defocusing, arising from cosmic voids [67]. When combined with measurements of the more usual gravitational lensing, or focusing, caused by over dense regions, these techniques may eventually provide another means whereby the validity of the HDIE model can be tested.

## 6. Summary

The location of the HDIE data point in Fig.2 is determined directly, and solely, by the measured integrated stellar density data of Fig. 1, while the HDIE limits are set by the ±1σ limits of that data fit to those power laws. The close agreement of HDIE with *Planck* dataset combinations, and the ease with which HDIE can account for today's dark energy density to resolve the cosmological coincidence problem, argues that HDIE should be considered a viable dynamic explanation for dark energy. Table 1. shows that HDIE compares well against the two main dark energy explanations, and an HDIE explanation would then enable the cosmological constant to take the more likely zero value.

While HDIE accounts for today's dark energy density value (requirement b above) without applying the Holographic principle, HDIE can only satisfy the other requirements (a, c and d) by utilizing that principle. Then, should HDIE be eventually proven to be the correct explanation for dark energy, it would provide very strong support for the Holographic Principle.